\documentclass[twocolumn,secnumarabic,amssymb,amsfonts,nobibnotes,aps,jcp,showpacs]
{revtex4-1}

\usepackage{amsmath}
\usepackage{latexsym}
\usepackage {mathrsfs}
\usepackage{float}
\usepackage{amssymb}
\usepackage{amsfonts}
\usepackage{graphicx}
\usepackage{textcomp}
\usepackage{hyperref}
\usepackage {mathrsfs}
\usepackage {pifont}
 1

\begin{document}

\title{Dynamics and Growth of Droplets Close 
to the Two-Phase Coexistence Curve in Fluids}
\author{Sutapa Roy and Subir K. Das$^*$}
\affiliation{Theoretical Sciences Unit, Jawaharlal Nehru Centre for Advanced 
Scientific Research, Jakkur P.O, Bangalore 560064, India}
\date{\today}

\begin{abstract}
Results from the state-of-the-art molecular dynamics simulations 
are presented for both equilibrium and nonequilibrium dynamics 
following vapor-liquid transition in a single component 
Lennard-Jones system. We have fixed the overall density close to 
the vapor-branch of the coexistence curve so that the liquid 
phase forms droplet structure in the background of vapor phase. 
In the equilibrium case, the motion of a single droplet is 
studied in both microcanonical and canonical ensembles, in the 
latter case a hydrodynamics preserving Nos\'{e}-Hoover thermostat 
was used to control the temperature. The droplet nucleation, motion, 
collision and coalescence dynamics in the nonequilibrium case 
were studied in the canonical ensemble with Nos\'{e}-Hoover 
thermostat. There it was observed 
that the average droplet volume grows linearly with time. 
Between two successive collisions, the size of the droplets 
remain same even though all the constituent particles do not 
move with the droplets \textminus~some leave, others join. 
It is seen that the number of original particles in a droplet 
decays exponentially fast. Results from a liquid-liquid 
transition are also presented in the equilibrium context. 
Dynamics of droplets in equilibrium appears to be at variant with 
the nonequilibrium case. 
\end{abstract}
\maketitle

\section{Introduction}\label{introduction}
\par
\hspace{0.2cm}The subject of nucleation remains interesting 
over many decades \cite{zettlemoyer,abraham,binder5,
kashchiev,binder1}. Physics of nucleation and 
subsequent growth is certainly challenging and answers to 
many fundamental questions in this branch are subjects 
of debate. On the other hand, understanding of these 
phenomena is crucial to many practical applications 
\cite{gelb,sing,squires1} including nanoscience and 
technology. 

\par
\hspace{0.2cm}
When a homogeneous system is quenched (say, by sudden change 
of temperature ($T$)) inside the coexistence curve 
the system phase separates into particle rich and particle 
poor domains \cite{bray,onuki,binderbook,jones}. 
While for quenches close to the critical 
density (or composition in a binary system) the phase 
separation or domain coarsening is spontaneous (referred to 
as spinodal decomposition), for ones close to the 
coexistence curve the system may reside in a metastable 
state in which case phase separation starts only after 
nucleation of droplets following rare long wave length 
fluctuations \cite{binder5}. 
\par
\hspace{0.2cm}
In this work, staying outside the so-called spinodal 
regime, we address 
the issue of dynamics of fluid droplets and their growth, 
once they are formed. 
Results are presented for both equilibrium and nonequilibrium 
situations, for a vapor-liquid as well as a liquid-liquid 
transition. For the liquid-liquid case, however, because 
of the computational demand, we will confine ourselves to 
studies in equilibrium condition. The primary emphasis 
in this paper being 
on the nonequilibrium 
dynamics \cite{bray,onuki,binderbook,jones}, 
below we briefly review the subject to place our work in 
appropriate perspectives. Note here that we expect the discussions 
provided below to be valid for both 
liquid-liquid and vapor-liquid transitions, even though some 
of them were originally meant for liquid-liquid cases or 
binary mixtures in general.
\par
\hspace{0.2cm}Coarsening of domains in solid binary mixtures is 
due to diffusive transport and follows the celebrated 
Lifshitz-Slyozov (LS) law \cite{lifshitz,bray,suman1,suman2}, 
where the average length, $\ell(t)$, of domains grows with 
time ($t$) as $t^{1/3}$. This scaling law is true 
irrespective of a quench close to the critical composition or 
near the coexistence curve. On the other hand, in fluids, 
hydrodynamics plays important role \cite{siggia,hansen,shaista1,
suman3,furukawa1,furukawa2}. There, for quenches close to the 
critical density or composition, where one has an 
interconnected domain structure, one expects three different 
regimes of growth. The early time diffusive LS mechanism is 
followed by a linear viscous 
\cite{siggia,bray,onuki,shaista1,suman3,
furukawa1,furukawa2} hydrodynamic regime with an 
exponent $\alpha=1$ and further by an 
inertial \cite{bray,onuki,furukawa1,furukawa2} 
hydrodynamic regime with an exponent $2/3$. It is 
argued that through the tube-like interconnected domains 
material flow becomes faster due to hydrodynamics. On the 
other hand, one has a disconnected droplet morphology close 
to the coexistence curve where the domains are expected to 
grow via collisions \cite{binder2,binder3,siggia,tanaka1,
tanaka2,tanaka3,kumaran1,kumaran2,roy1,perrot}.
\par
\hspace{0.2cm}There are different possibilities for the 
mechanism of the above mentioned collision events. If there 
exists any inter-droplet force \cite{tanaka1,tanaka2,kumaran1,
kumaran2}, the dynamics of droplets is going to be 
deterministic. Other possibility is that the droplets exhibit 
Brownian motion and collide randomly 
\cite{binder2,binder3,siggia}. Even though both the cases are 
discussed in the literature, the dynamics of coarsening of 
liquid droplets received much less attention as opposed to 
spinodal decomposition. In this work our primary focus is 
on the late stages of droplet growth and understanding the 
microscopic details of the motion of the droplets. Some 
preliminary results in this context were published in a recent 
communication \cite{roy1}.  
\par
\hspace{0.2cm}According to Binder and Stauffer (BS) 
\cite{binder2,binder3,siggia}, if the motion of the 
droplets is Brownian, $\ell(t)$ should grow as 
$t^{1/3}$. In that case, one can write the equation for the 
rate of change of droplet density $n$ as 
\begin {eqnarray}\label{BS}
\frac{dn}{dt}={C_1}{D_d}{\ell}{n^2},
\end{eqnarray}
where $D_d$ is the droplet diffusivity 
and $C_1$ is a constant. Using $n \propto {1/\ell^3}$ and 
noting that, according to Stokes-Einstein-Sutherland (SES) 
relation \cite{hansen,squires2,das2}, ${D_d}\ell$ is a constant, 
solution of Eq.(\ref{BS}) gives
\begin {eqnarray}\label{powerlaw}
\ell(t)\sim t^{1/3}.
\end{eqnarray}
\par
\hspace{0.2cm}Tanaka proposed that based on the density 
of droplets, the motion could be deterministic 
\cite{tanaka1,tanaka2}. He argued, the BS mechanism 
will be valid only in the low droplet density limit, 
for minority species (in a binary mixture context) 
volume fraction $\phi < 0.06$. Then the interdroplet 
distance is much larger than $\ell$ such that the 
droplet-droplet interaction can be neglected. In such 
situations the droplets can freely undergo Brownian 
motion and random collisions. On the other hand, at high 
droplet density (with $\phi>0.06$), the diffusion 
field or the concentration gradients around neighbouring 
droplets get coupled which gives rise to an attractive 
interaction between the droplets \cite{tanaka1,tanaka2,tanaka3}. 
This attractive force eventually induces a direct collision. 
This can also be appreciated by considering 
the excess free energy $\sim (\nabla \psi)^2$ coming from 
the concentration (or density) gradient, $\psi$ being the 
concentration. The excess energy provides a drive for 
removing the gradient zone, thus causing the 
attractive interaction between the neighbouring droplets. 
Note that this gradient term gives rise to the diffusion 
under concentration gradient in Ginzburg-Landau formalism 
\cite{bray,onuki,jones}. 
This sort of interdroplet interaction will surely accelerate 
the droplet growth mechanism. However, for the form of 
interaction obtained by Tanaka, 
this deterministic motion of droplets provides the exponent 
for the growth law same as the Brownian mechanism but 
a higher amplitude which, for the sake of brevity, we do 
not discuss here. The brief discussion below, however, 
provides idea of how to obtain some knowledge about it from 
computer simulations. 
\par
\hspace{0.2cm} 
The ratio of the growth amplitudes $A_{LS}$ and $A_{BS}$ 
in the LS and BS mechanisms has been pointed out to be 
\cite{siggia,tanaka1,tanaka2}
\begin {eqnarray}\label{A_ratio}
A_{BS}/A_{LS}=K\phi^{1/3};~K\simeq 6.
\end{eqnarray}
It has 
also been argued \cite{tanaka1,tanaka2} that 
$K \simeq 4.84$ due to some 
ambiguity in the value of constant $C_1$ in Eq. (\ref{BS}). 
By incorporating appropriate dynamics, in computer 
simulations, it is possible to estimate this ratio. This, 
in turn, will provide the realization of actual 
mechanism, Brownian or non-Brownian. We will use this fact 
plus other means to figure out the presence of inter-droplet 
interaction.
\par
\hspace{0.2cm} 
In this work, in addition to 
estimating the exponent of the fluid droplet growth, 
we investigate the true mechanism for the growth. 
Further, we aim to gain knowledge about the microscopic 
origin of droplet motion in both equilibrium and 
nonequilibrium contexts, validity of SES relation for 
these objects, etc. Number of results we have presented here 
and methodologies used for that purpose are nonconventional 
in the context of coarsening dynamics.
\par
\hspace{0.2cm}The rest of the paper is organized as 
follows. In Sec. II we discuss the model and the 
methodologies. Results are presented in Sec. III. Finally, 
Sec. IV concludes the paper with an outlook for 
some of the future possibilities.

\section{Models and Methods}\label{model}
\par
\hspace{0.2cm}For this study, we choose a model in which 
particles of equal mass ($m$), at distance $r$ apart, 
interact via \cite{roy1,suman3}
\begin {eqnarray}\label{LJ1}
u(r=|{{\vec r}_i}-{{\vec r}_j}|)=U(r)-U(r_c)-(r-r_c)
({{dU}/{dr}})_{{r}=r_c},~~~
\end{eqnarray}
where, for the vapor-liquid transition, 
\begin {eqnarray}\label{LJ2}
U(r)=4\varepsilon\big[({\sigma}/{r_{ij}})^{12}-
({\sigma}/{r_{ij}})^6\big]
\end{eqnarray}
is the standard Lennard-Jones (LJ) pair potential, 
$r_c$ is a cut-off distance for the LJ potential 
(introduced for faster computation), $\sigma$ is the 
particle diameter and $\varepsilon$ is the interaction 
strength. For the present work, we have set $r_c=2.5\sigma$. 
The model exhibits a gas-liquid phase transition \cite{roy1} 
with 
critical temperature $k_BT_c\simeq0.9\varepsilon$ and 
critical density $\rho_c\simeq0.3$. Note 
that $\rho$ is defined as $N\sigma^3/L^3$ where $N$ is 
the number of particles in a cubic box of linear 
dimension $L$ in units of $\sigma$. For the rest of the 
paper, we set $\varepsilon=1$, $m=1$, $\sigma=1$ and the 
Boltzmann constant $k_B=1$. This sets the LJ time unit 
$\sqrt{m\sigma^2/\varepsilon}=1$. 
\par
\hspace{0.2cm} The nonequilibrium dynamics using this 
model is studied via the molecular dynamics (MD) 
\cite{frenkel,allen} simulations in NVT ensemble using two 
different thermostats to control the temperature. 
To probe the hydrodynamic effects we have used the 
Nos\'{e}-Hoover thermostat (NHT) \cite{frenkel,hoover}. 
On the otherhand, an Andersen thermostat (AT) 
\cite{frenkel} was used 
to study the difference between a growth via the 
droplet diffusion mechanism, studied via NHT, and the 
simple LS mechanism. We have also used the perfectly 
hydrodynamics preserving NVE ensemble to probe the 
dynamics of a single droplet in the equilibrium 
situation. Even though this latter ensemble preserves 
hydrodynamics perfectly, it cannot be used for the 
nonequilibrium study since it cannot keep the 
temperature constant during the coarsening process.
\par
\hspace{0.2cm}
The time step for the 
MD runs was chosen to be $\Delta t=0.005$ in the 
Verlet velocity algorithm \cite{frenkel}. We applied 
periodic boundary conditions in all directions. All 
results correspond to $T=0.6$ and $\rho=0.05$. Note 
that at this temperature the coexistence vapor density 
is $\simeq 0.01$. Unless otherwise mentioned, all 
results for the nonequilibrium case, obtained using 
the NHT, are averaged over $12$ independent initial 
configurations, with $L=100$. 
\par
\hspace{0.2cm} At $t=0$, initial configurations with 
homogeneous density, prepared at a high temperature 
$T=5$, were quenched to $T=0.6$ inside the coexistence 
curve. To understand the subsequent domain growth and 
related dynamics, we computed several important 
quantities which we discuss below. 
\begin{figure}[htb]
\centering
\includegraphics*[width=0.4\textwidth]{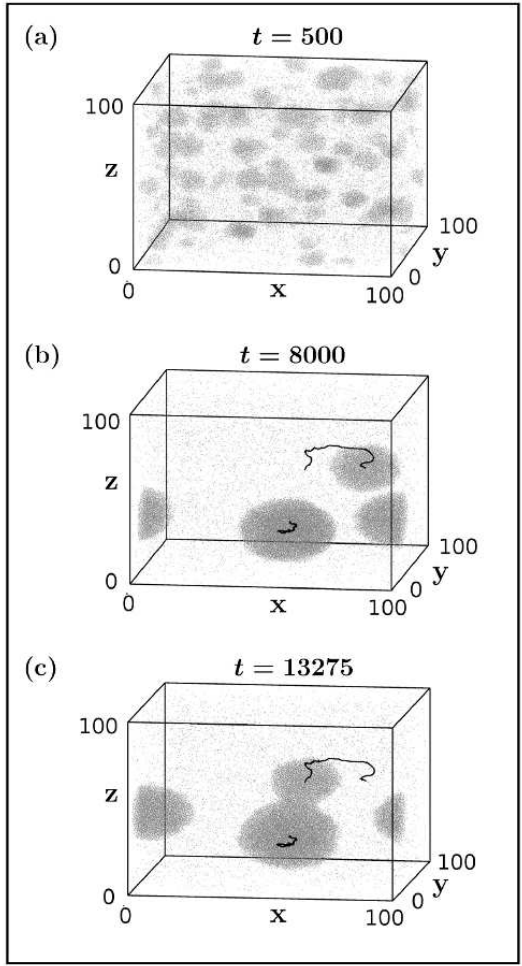}
\caption{\label{fig1}Evolution snapshots following a 
temperature quench of a low density ($\rho=0.05$) 
homogeneous fluid, prepared at a high temperature, to a 
temperature $T=0.6$, inside the coexistence curve. The 
zig-zag lines in (b) and (c) are the trajectories of 
two droplets during the interval between $t=8000$ 
and $t=13275$.}
\end{figure}
\begin{figure}[htb]
\centering
\includegraphics*[width=0.4\textwidth]{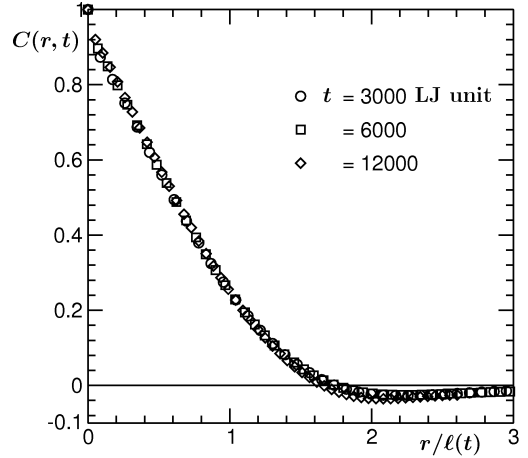}
\caption{\label{fig2}Scaling plot of the correlation 
function $C(r,t)$ as a function of $r/\ell(t)$. Data from 
three different times are used. The values of $\ell(t)$ 
were obtained from the decay of $C(r,t)$ to $1/4$th its 
maximum value. The results correspond to $\rho=0.05$, 
$L=100$ and $T=0.6$ and an averaging over $12$ independent 
initial configurations.}
\end{figure}
\begin{figure}[htb]
\centering
\includegraphics*[width=0.4\textwidth]{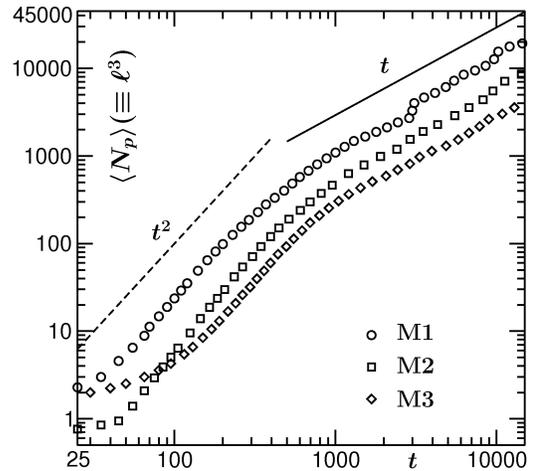}
\caption{\label{fig3}Log-log plots of the average number of 
particles in droplets, $\langle N_p \rangle$, or the average 
droplet volume, $\ell^3$, as a function of time, during 
the nonequilibrium evolution, calculated via three 
different methods. Possibilities for various different 
power laws are indicated.}
\end{figure}

\begin{figure}[htb]
\centering
\includegraphics*[width=0.4\textwidth]{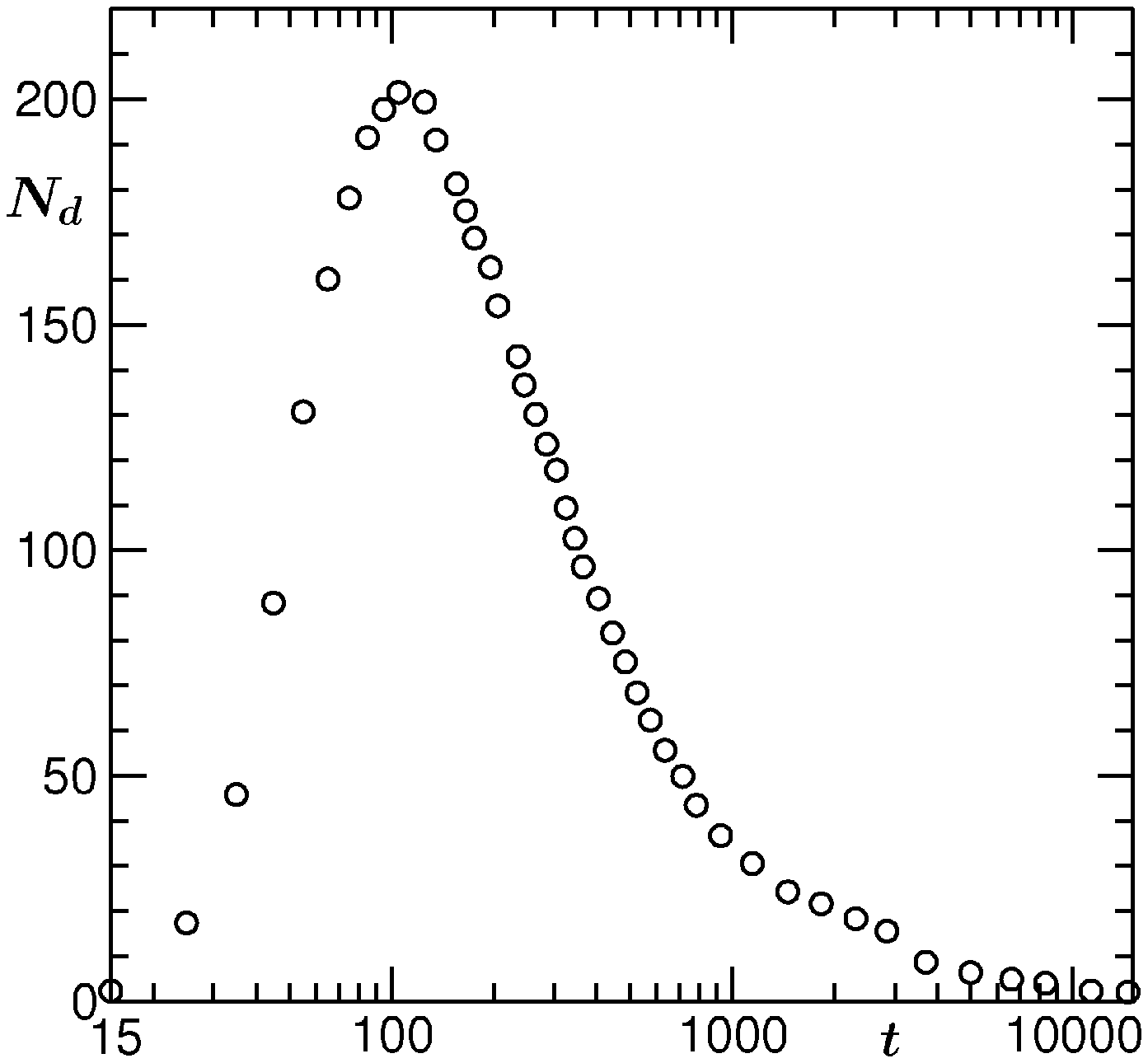}
\caption{\label{fig4}Semi-log plot of the average number 
of droplets, $N_d$, as a function of time. All the system 
parameters are same as the previous figures.}
\end{figure}
\par
\hspace{0.2cm}
The average domain size, $\ell(t)$, was calculated 
using the following scaling property of the two-point 
equal time order-parameter correlation function, 
$C(r,t)$, ($r$ being the spatial separation between 
two points) \cite{bray}
\begin {eqnarray}\label{scor}
C(r,t)\equiv \tilde {C}(r/\ell(t)).
\end{eqnarray}
To facilitate the calculation of $C(r,t)$, we have 
mapped \cite{suman1,suman2,daspuri} a continuum fluid system 
into a simple cubic 
lattice of lattice constant $\sigma$, where particles 
in the actual system were moved to the nearest lattice 
sites. In the mapped system, a site, with index $i$, 
occupied by a particle was assigned a spin value 
$S_i=+1$, otherwise $-1$. Note that in this spin 
language the value of $\phi$, the volume fraction of the 
minority species (in this case the up spins), appears to 
be $0.054$. Finally, $C(r,t)$ was calculated from the 
standard formula $C(r,t)=\langle{{S_i}{S_j}}\rangle - 
\langle {S_i}\rangle \langle {S_j}\rangle$, 
$r=|{\vec i}-{\vec j}|$. Further, $\ell(t)$ was 
also calculated directly by sweeping through the 
lattice and identifying the domain boundaries. 
Another important quantity, may be difficult to 
calculate, that gives information about $\ell(t)$ 
is the number of particles in droplets, $N_p$, since
$\langle N_p(t)\rangle  \propto \ell(t)^3$. 
The distribution function of $N_p$ should obey the 
scaling relation
\begin {eqnarray}\label{npscaling}
P(N_p,t)\equiv {\frac{1} {\langle N_p \rangle}}
\tilde P({N_p/\langle N_p \rangle}).
\end{eqnarray}
In Eqs. (\ref{scor}) and (\ref{npscaling}), 
$\tilde {C}$ and $\tilde P$ are master functions 
independent of time.
\begin{figure}[htb]
\centering
\includegraphics*[width=0.405\textwidth]{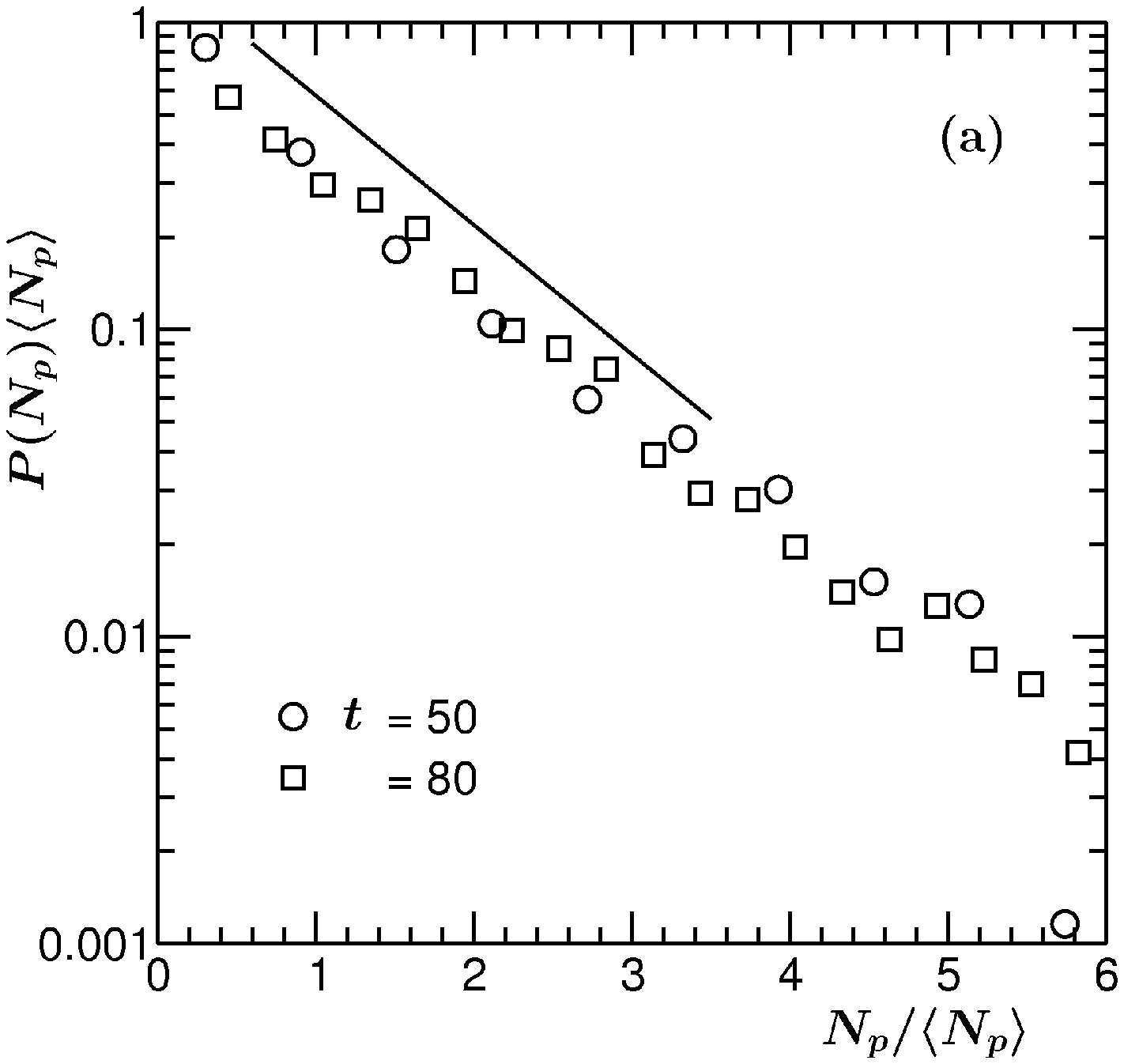}
\vskip 0.5cm
\includegraphics*[width=0.4\textwidth]{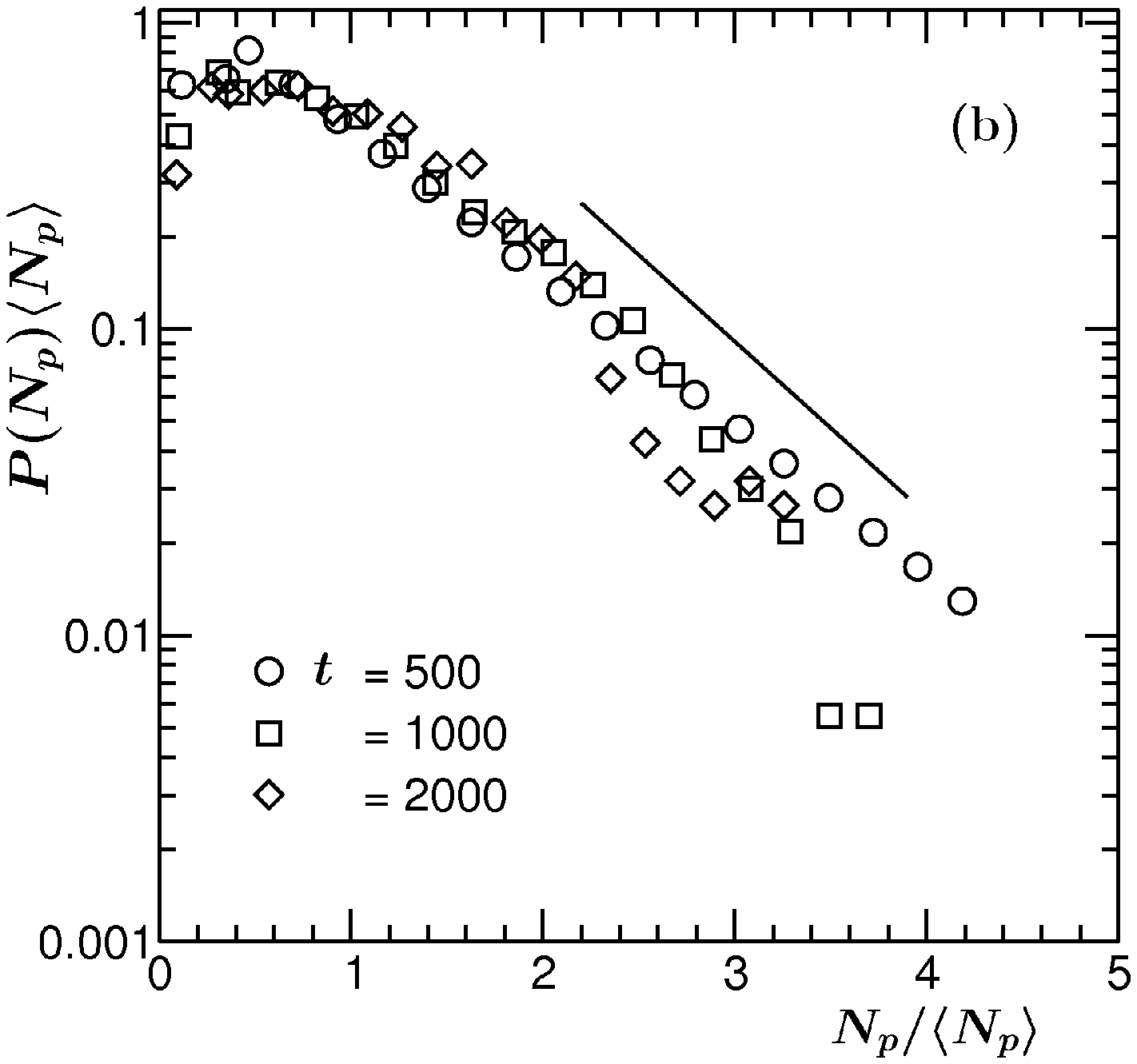}
\caption{\label{fig5}Semi-log plots of the scaled distribution 
functions of the number of particles in droplets ($N_p$), from 
different times. The solid lines corresponds to exponential 
decay. (a) During droplet nucleation stage. (b) During droplet 
growth stage.}
\end{figure}
\begin{figure}[htb]
\centering
\includegraphics*[width=0.39\textwidth]{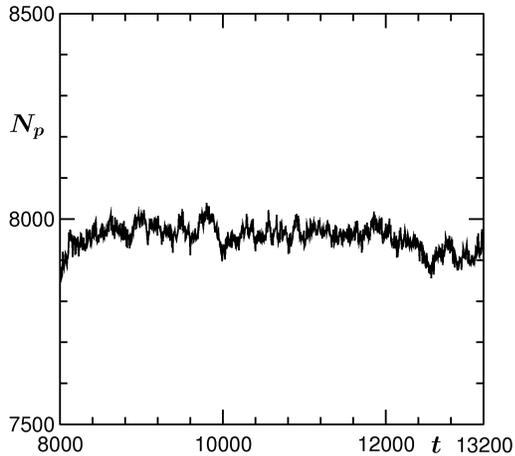}
\caption{\label{fig6}Plot of the number of particles 
in a typical droplet before it undergoes a collision.}
\end{figure}

\begin{figure}[htb]
\centering
\includegraphics*[width=0.38\textwidth]{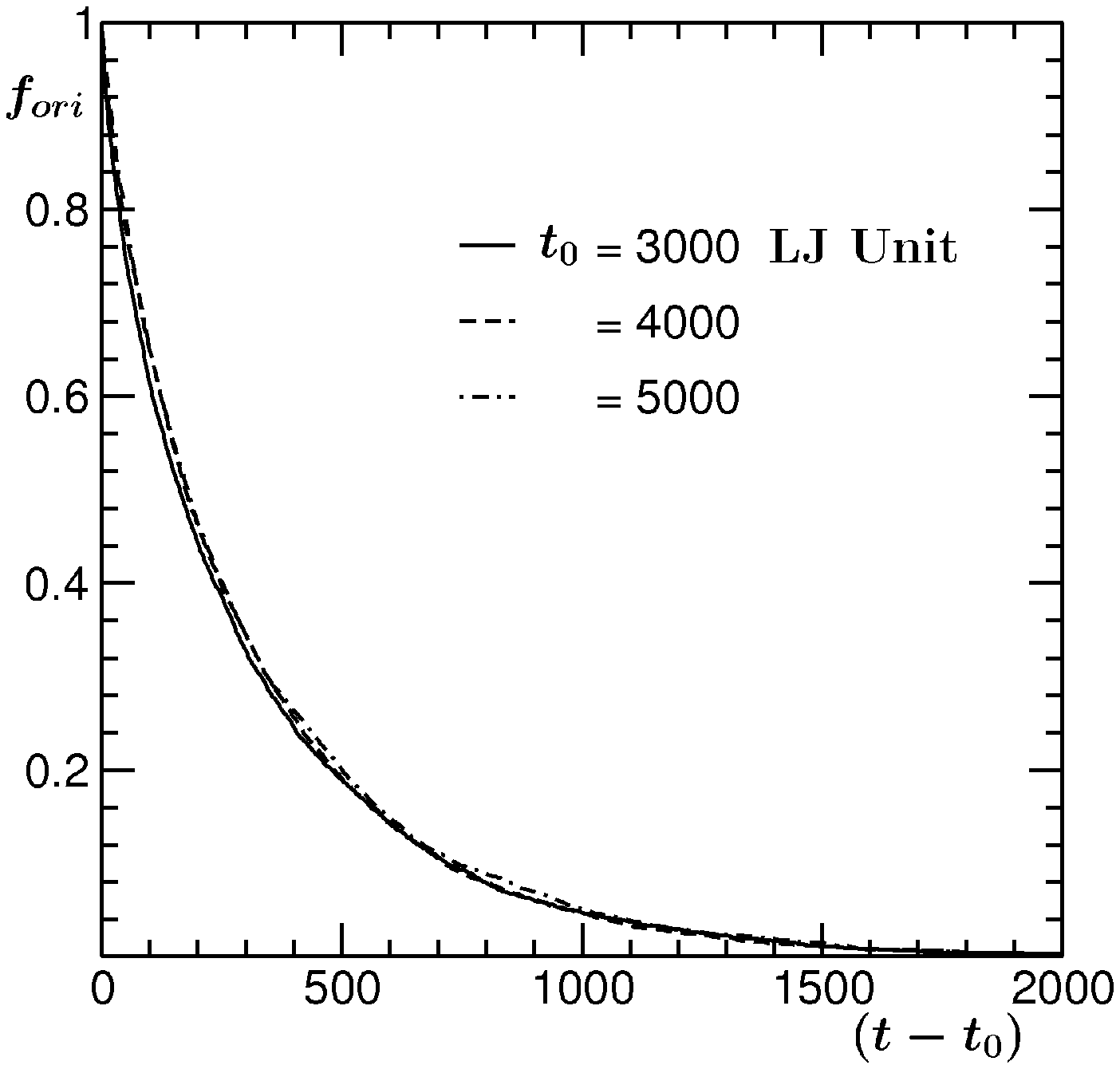}
\caption{\label{fig7}Plots of the fraction, $f_{ori}$, of 
original particles at time $t_0$, that move with the droplet, 
as a function of $(t-t_0)$. Results for three different values 
of $t_0$ are shown.}
\end{figure}
\par
\hspace{0.2cm} To calculate $N_p$, one first needs 
to identify the droplets for which we used the 
following method. We calculated the density 
surrounding each particle, by defining a neighbourhood of radius 
$R$, using the continuum 
configuration and if the density was higher than the 
critical value, the particle was marked as a constituent 
of any of the droplets. Note that this radius cannot be too 
small or too large. In the limiting case $R=$ the particle 
radius, one is not including any neighbour. On the otherhand, for 
large value of $R$, there is the risk of moving to a 
different phase if the central particle is close enough to the 
interface. We have taken care of these facts. 
Next, depending on the spatial 
separations between these marked particles, different 
droplets were identified and the particles were indexed 
according to their parent droplets. Following this, 
information about $N_p$, droplet diameter ($d_d$), total 
number of droplets ($N_d$) in the system at time $t$, 
etc., can be obtained in a straight forward manner. 
Note that to obtain exact information about $d_d$ 
(most accurate measure of $\ell$), we need to calculate 
the density inside the droplet also. 
\par
\hspace{0.2cm}
In order to study the nature of droplet motion, we 
calculated the mean squared displacement (MSD) of the 
centre of mass (CM) of a droplet as \cite{hansen}
\begin {eqnarray}\label{msd}
MSD=\langle(\vec R_{CM}(t)-\vec R_{CM}(0))^2\rangle,
\end{eqnarray}
where the position vector of the droplet CM, 
$\vec R_{CM}$, has the standard definition
\begin {eqnarray}\label{rcm}
\vec R_{CM}(t)={\frac {1}{N_p}}{\sum_{i=1}^{N_p}\vec r_i(t)}. 
\end{eqnarray} 
According 
to our observation, as the 
droplets move , the constituent particles also keep 
changing \textminus~some leave, some join. Therefore, 
while calculating the MSD, at each instant particles within 
the droplet were identified and $\vec R_{CM}$ at that instant 
was calculated using those particles only. The diffusivity, 
$D_d$, of a droplet was calculated using the relation 
\cite{hansen}
\begin {eqnarray}\label{diffusivity}
D_d(t)=\lim_{t \rightarrow \infty} 
{\frac{MSD}{6t}},
\end{eqnarray}
and verified if $D_d$ follows the SES relation 
\cite{hansen,squires2,das2}
\begin {eqnarray}\label{ses}
D_d \sim \frac {T}{6\pi\eta {d_d}},
\end{eqnarray} 
$\eta$ being the effective shear viscosity of the background 
fluid. As we will see, it appears that this holds well in 
the equilibrium situation for both vapor-liquid and 
liquid-liquid transitions. For the latter, we briefly discuss 
the model below.

\par
\hspace{0.2cm} 
In the liquid-liquid $(A+B)$ case, we choose a symmetric 
model with the same interaction (\ref{LJ1}) but the LJ 
potential given as
\begin {eqnarray}\label{binaryLJ1}
U(r=|{{\vec r}_i}-{{\vec r}_j}|)=4\varepsilon_{ij}
\Big\lbrack(\frac{\sigma_{ij}}{r_{ij}})^{12}-
(\frac{\sigma_{ij}}{r_{ij}})^6\Big\rbrack,
\end {eqnarray}
with 
\begin {eqnarray}\label{sigma}
\sigma_{AA}=\sigma_{BB}=\sigma_{AB}=\sigma,\nonumber
\end {eqnarray}
\begin {eqnarray}\label{epsilon}
\varepsilon_{AA}=\varepsilon_{BB}=2\varepsilon_{AB}=\varepsilon.
\end {eqnarray}
We set the overall density of the system to unity. 
More details on this model and related 
phase diagram can be found in Refs. \cite{das1,roy2}. 
We just mention here that this binary mixture has 
$T_c \simeq 1.42$ and because of the symmetry in Eq. (\ref{epsilon}), 
the critical composition is $50\%$ $A$ and $50\%$ $B$. 
In this work, the concentration 
of $A$ species, $x_A(=\frac{N_A}{N}$, $N_A$ and $N$ being 
respectively the number of $A$ particles and total number 
of particles$)$, was set to $0.15$ and temperature to $T=1.0$. 
For this choice of parameters, droplets of $A$ species are 
formed in the background of $B$. All results for the binary 
fluid are obtained from $8$ independent initial 
configurations in NVE as well as NVT (NHT) ensembles. Periodic 
boundary conditions were applied in all directions. In this 
case the droplets were identified following a similar 
procedure as in the vapor-liquid case, considering the $A$ 
particles only. Subsequently the droplet MSD and 
droplet diffusivity were calculated using Eq. (\ref {msd}) and 
Eq. (\ref{diffusivity}) respectively.
\par
\hspace{0.2cm} Next we move on to present the results. 
Unless otherwise mentioned, all results correspond to the 
vapor-liquid transition.

\section{Results}\label{results}
\par
\hspace{0.2cm}In Fig. \ref{fig1}, we show the snapshots 
during the nonequilibrium evolution of a vapor-liquid 
system at $T=0.6$, starting from a homogeneous density, 
prepared at a high temperature. Already at $t=500$, 
there are reasonably well formed droplets in the system. 
The other two snapshots in (b) and (c) demonstrate the 
growth of the droplets with time. In (b) and (c) we have 
also drawn the trajectories of two droplets starting from 
$t=8000$ to $13275$. Note that the droplets collide with each 
other at $t=13275$. While the trajectories certainly have 
some degree of randomness, one needs further probes to 
confirm whether they are truly Brownian or not. A look at 
the mean squared displacements of the droplets should 
prove useful, to which we will come back later.
\par
\hspace{0.2cm}In Fig. \ref{fig2} we plot $C(r,t)$ 
vs the scaled distance $r/\ell(t)$ for three different 
times. Note that $\ell(t)$ used here was obtained from the 
decay of the $C(r,t)$ to $1/4$th its maximum value. The 
excellent quality of data collapse indicates 
self-similarity of the structures in this time regime. 
\par
\hspace{0.2cm}
Fig. \ref{fig3} shows the plots of the average number of 
particles in droplets, $\langle N_p\rangle$, and the average 
droplet volume, $\ell(t)^3$, vs $t$, $\ell(t)$ being 
calculated via different methods. Double-log scale 
is used. Note that $\langle N_p\rangle \propto \ell(t)^3$ 
holds under the assumption that the density 
in the liquid droplets does not change over time. 
We calculated $N_p$ directly from the knowledge of the 
constituent particles of droplets (M1), $\ell(t)$ from 
the decay of the correlation functions (M2) and by the 
method of sweeping through the lattice and identifying 
the domain interfaces (M3). All these were 
discussed in the previous section. The assumption of constant 
density of droplets does not hold at early time. Because of 
this there is some qualitative discrepancy of measure M1 with 
the other ones at early time. The solid lines 
in this plot correspond to different possible power-law 
scaling regimes. Following an initial faster growing regime 
with an exponent $2/3$, the system enters into a regime 
where $\ell(t) \sim t^{1/3}$. This latter regime may be 
due to the Brownian droplet diffusion and collision mechanism 
of BS \cite{binder2}. Certainly the very nice consistency of 
all the three data sets with the $t^{1/3}$ behavior confirms 
the growth exponent predicted by BS, if not the mechanism. 
Alongside, it also validates our different methods of 
identifying $\ell$. A discussion of the possibility of a 
$t^{2/3}$ behavior, at early time, is provided 
in Ref.\cite{roy1}. Possibility of a linear growth in this 
regime can also not be ruled out. This is considering the 
fact that a crossover is never very sharp and what appears 
as $2/3$ may well be due to a gradual crossover from linear 
to $1/3$ law. Possibility of a linear behavior was discussed 
by Tanaka and not an objective of this paper. 
\begin{figure}[htb]
\centering
\includegraphics*[width=0.41\textwidth]{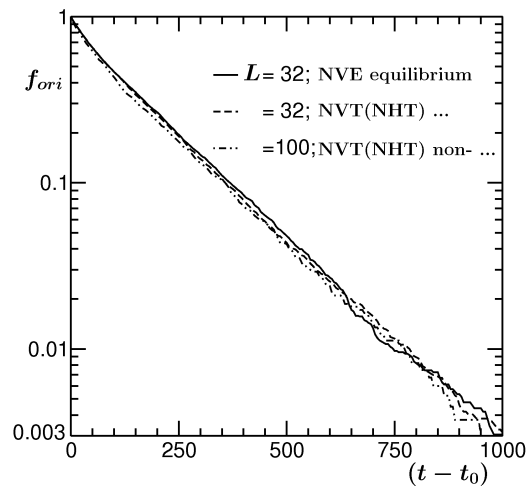}
\caption{\label{fig8}Plots of $f_{ori}$ as a 
function of $t-t_0$. Results are presented for three different 
cases: equilibrium situation (only one droplet present) in 
the NVE ensemble as well as with NHT and the nonequilibrium 
situation with the NHT. In all the cases the sizes of the 
droplets are nearly the same. A semi-log scale is used.}
\end{figure}

\begin{figure}[htb]
\centering
\includegraphics*[width=0.375\textwidth]{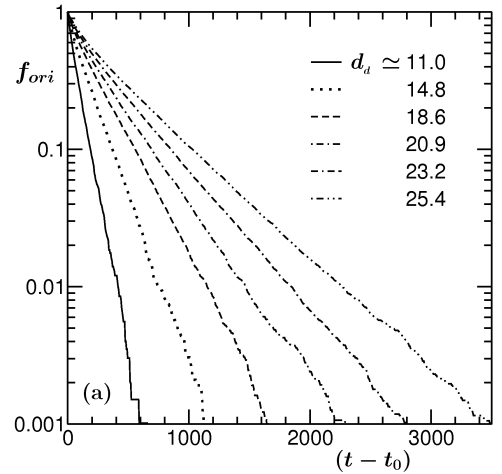}
\vskip 0.5cm
\includegraphics*[width=0.385\textwidth]{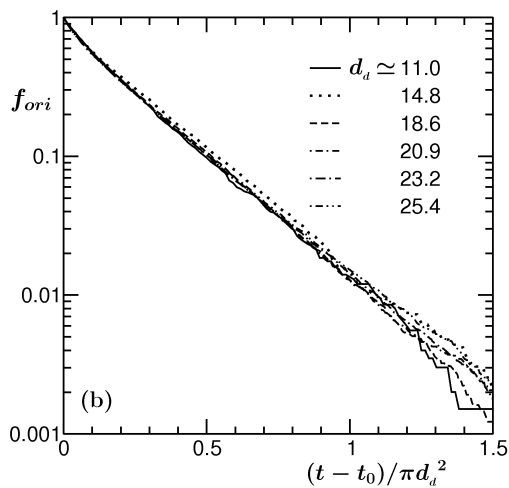}
\caption{\label{fig9}(a) Plots of $f_{ori}$ as a function 
of $t-t_0$ for different droplet sizes, obtained from 
equilibrium configurations using NVE ensemble. (b) Same as 
(a), but the $x$ axis scaled by $\pi {d_d}^2$.}
\end{figure}
\par
\hspace{0.2cm}In Fig. \ref{fig4} we plot the average number 
of droplets in the system, $N_d$, vs $t$, on a log-linear 
scale. As observed, beyond a time $t_g=t=115$, $N_d$ decays 
due to droplet coalescence (during which the 
average domain size increases, see Fig. \ref{fig3}). 
During time before $t_g$, mostly droplets keep forming. It 
will be interesting to study how in these two different 
regimes the probability distributions of $N_p$, $P(N_p,t)$, 
compare. In Figs. \ref{fig5}(a) and (b) we show the scaling 
plots of the distribution functions [see Eq. (\ref{npscaling})] 
using data from different times, as indicated in the figure, 
on a semi-log scale. Part (a) here is for 
$t<t_g$ and (b) for $t>t_g$. The solid lines there correspond 
to exponential decay. In the large droplet limit 
the distributions from both the time regimes are consistent 
with exponential decay, but difference exists in the small 
droplet size regime. Fast exponential decay in Fig. \ref{fig5}(a) 
from the very beginning is suggestive that all the droplets 
are roughly of same size which is expected at the formation 
stage. On the other hand, when the droplets are growing, 
there must be dispersion in size. This is reflected in 
Fig. \ref{fig5}(b) where for small values of the scaling 
variable $N_p/\langle N_p\rangle$, the distribution has a 
nearly flat character. This result is in agreement with 
previous studies \cite{suman2,suman3}, where instead of 
the domain volumes, distribution of domain lengths 
were studied.
\begin{figure}[htb]
\centering
\includegraphics*[width=0.38\textwidth]{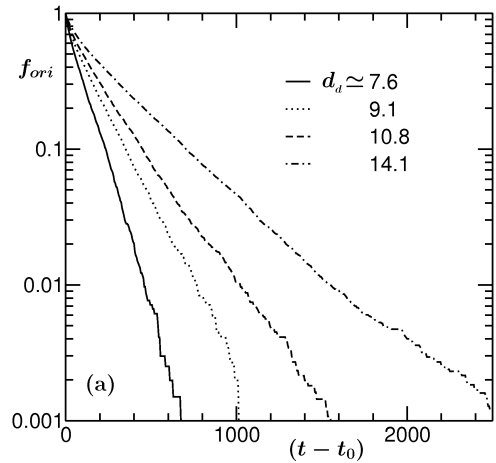}
\vskip 0.5cm
\includegraphics*[height=0.375\textwidth,width=0.38\textwidth]{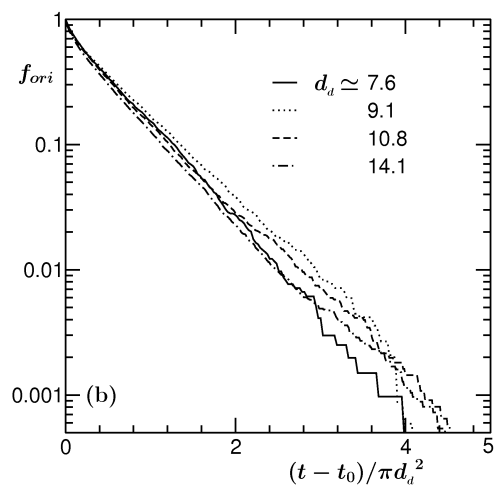}
\caption{\label{fig10} Same as Fig. \ref{fig9}, but in 
the binary fluid context.}
\end{figure}

\begin{figure}[htb]
\centering
\includegraphics*[width=0.38\textwidth]{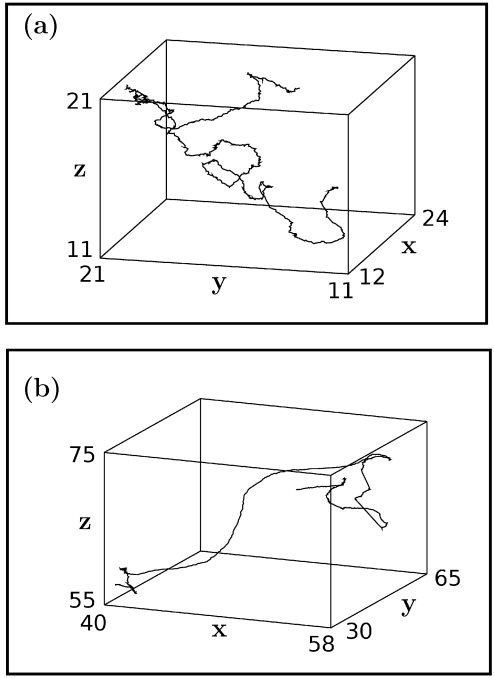}
\caption{\label{fig11}(a) Trajectory of a typical droplet 
the system is in equilibrium, i.e., there is 
only one droplet. The result was obtained using 
NHT. 
(b) Same as (a) but in the non-equilibrium situation. In 
both the cases, only parts of the boxes are shown.}
\end{figure}
\par
\hspace{0.2cm}Next we move on to discuss the droplet growth 
mechanism. We will be dealing only with the late stage 
when the density inside the droplets reached ``saturation". 
First, to rule out the possibility of LS mechanism, we 
provide the following discussion here. 
\begin{figure}[htb]
\centering
\includegraphics*[width=0.385\textwidth]{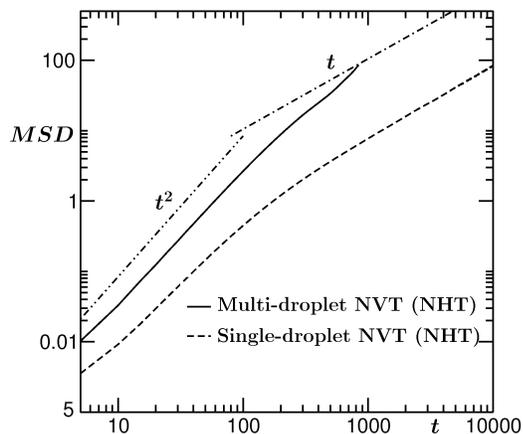}
\caption{\label{fig12}Log-log plots of the mean 
squared displacements vs time, of droplets in nonequilibrium 
(multi-droplet) and equilibrium (single-droplet) situations. 
Both correspond to an averaging over $6$ droplets, obtained 
using NVT (NHT). Note that in both the cases the droplets 
are of approximately same size.}
\end{figure}

\begin{figure}[htb]
\centering
\includegraphics*[width=0.37\textwidth]{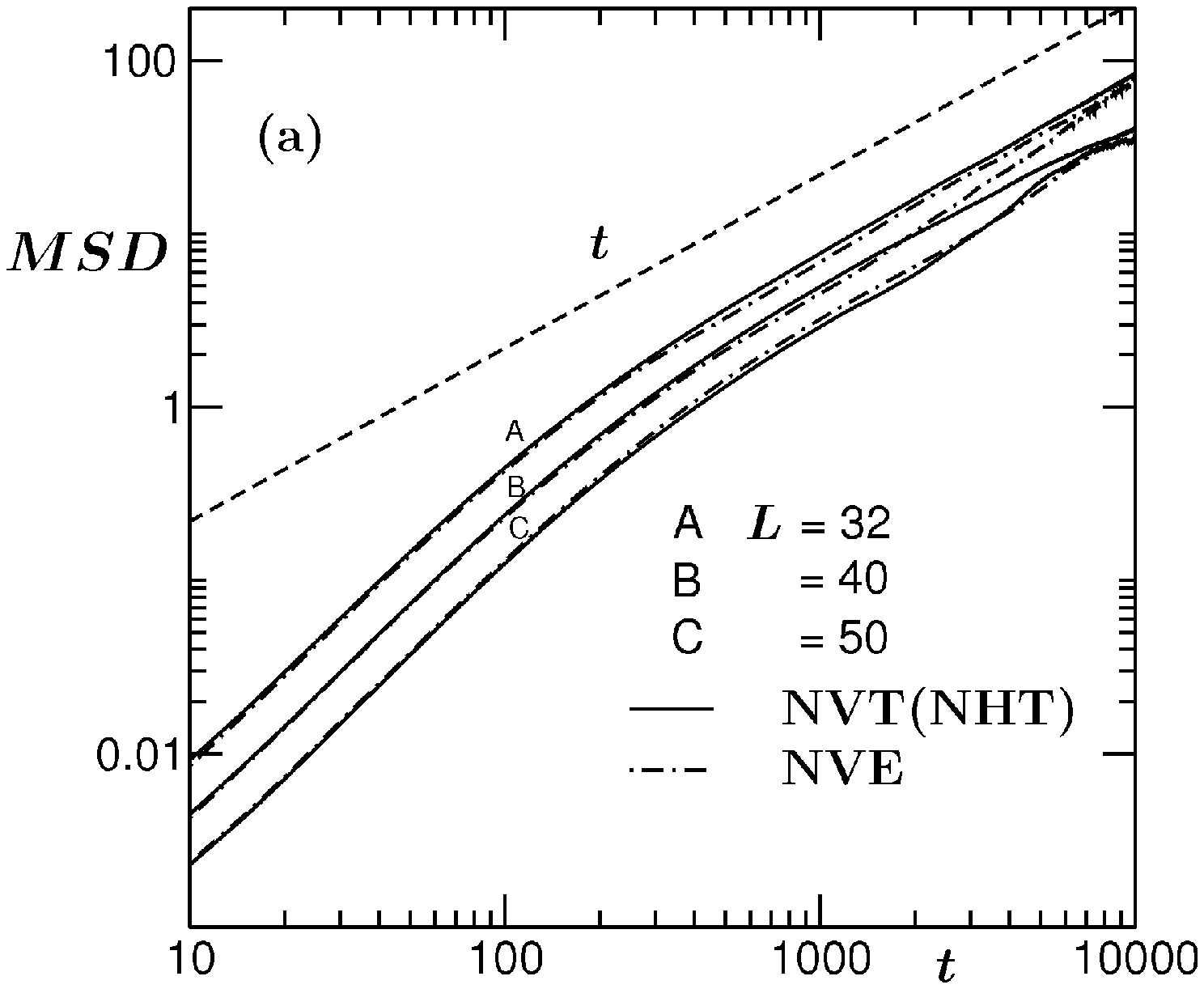}
\vskip 0.5cm
\includegraphics*[width=0.35\textwidth]{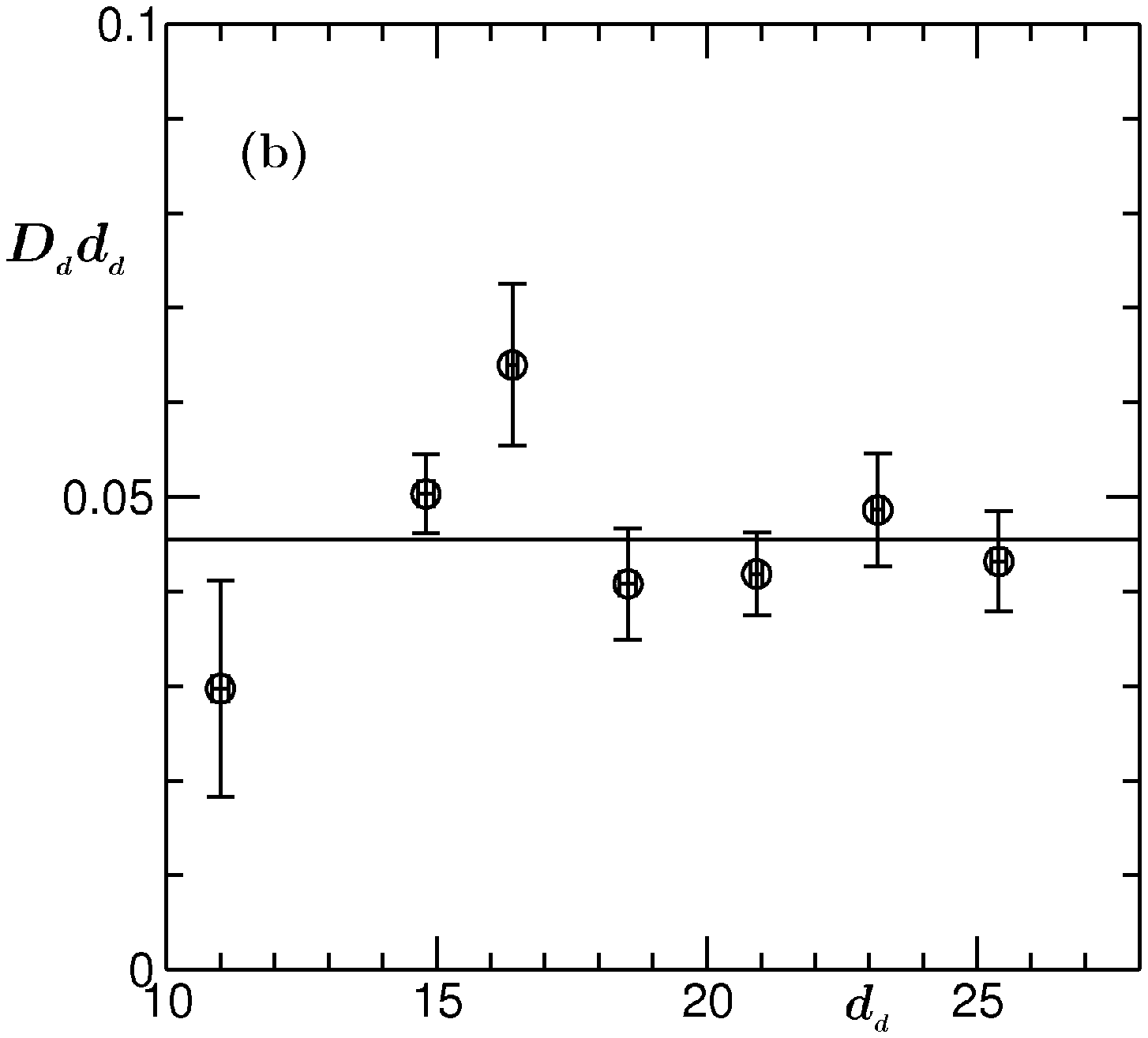}
\caption{\label{fig13}(a)Log-log plots of the mean squared 
displacements vs time, of droplets from equilibrium 
configurations, obtained using NVE as well as NVT (NHT). 
All results correspond to an averaging over 
$6$ droplets. 
(b) Plot of ${D_d}{d_d}$ vs the droplet size 
$d_d$, where $D_d$ is the droplet diffusivity. The results 
were obtained in microcanonical ensemble with an averaging 
over $6$ independent initial configurations.}
\end{figure}
\par
\hspace{0.2cm}As we mentioned in the introduction, 
the LS mechanism of domain growth also leads to an exponent 
$1/3$, same as BS. To distinguish between the two, 
we studied the droplet growth for the same overall density 
using AT also. AT being purely 
stochastic in nature will give rise to the LS mechanism only. 
For the sake of brevity we do not present the results here. 
In this case, it appears that the droplets are almost static 
which is consistent with the LS mechanism where domains 
grow primarily via the particle concentration diffusion. As we will 
see below, this is not the case when we use NHT.
\begin{figure}[htb]
\centering
\includegraphics*[width=0.37\textwidth]{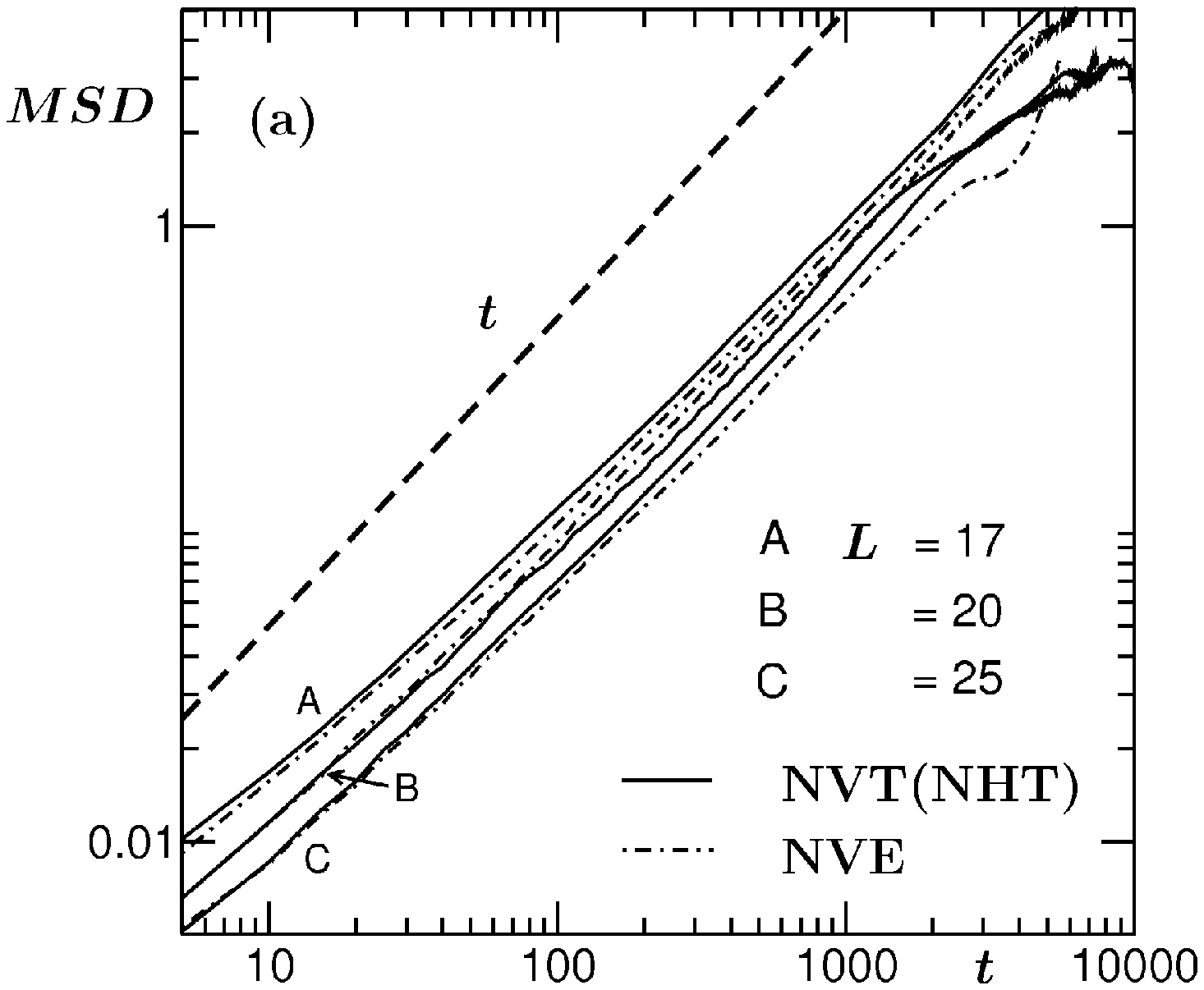}
\vskip 1.2cm
\includegraphics*[width=0.35\textwidth]{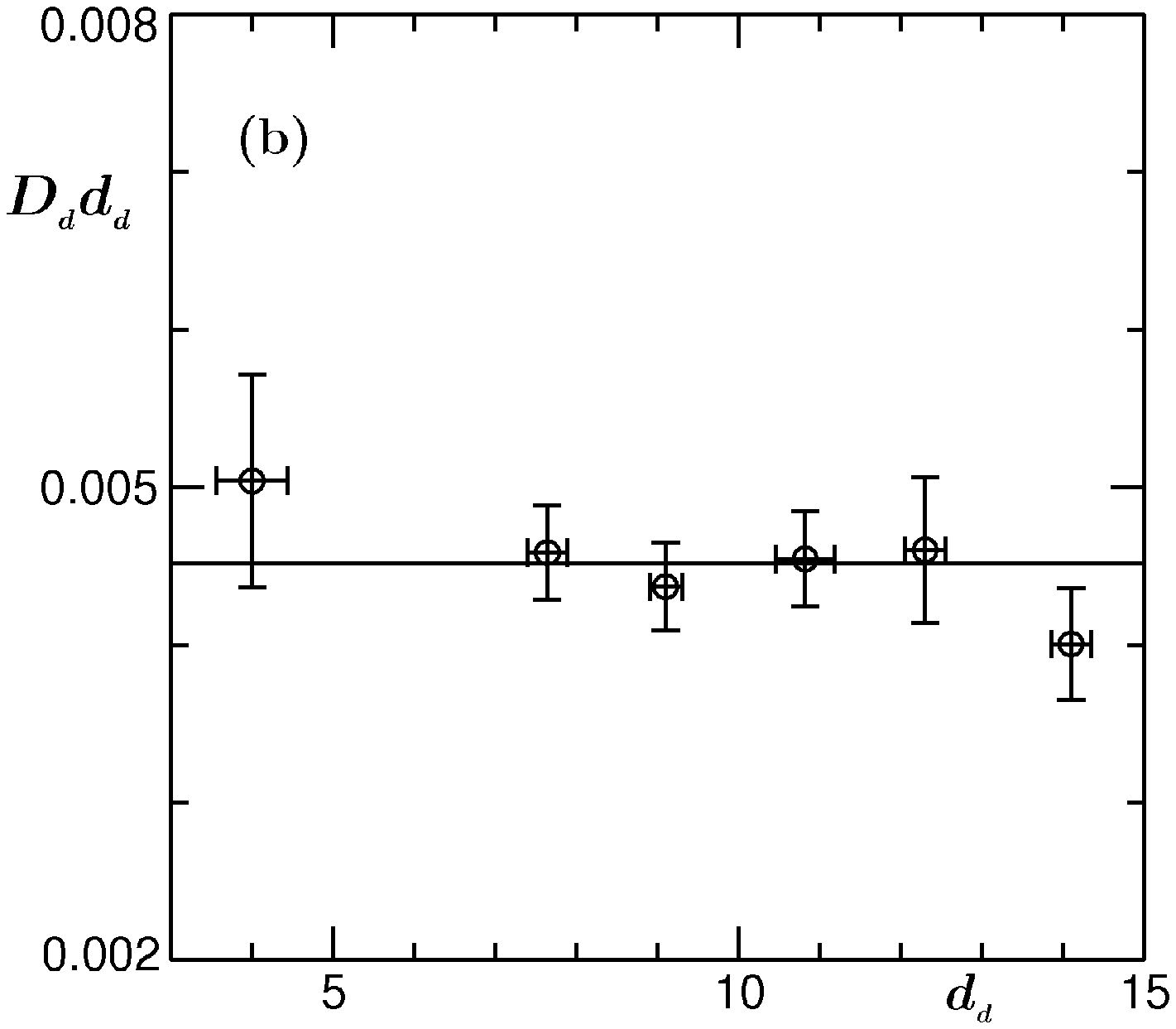}
\caption{\label{fig14}(a) Log-log plots of the mean squared 
displacements of equilibrium droplets, from the binary fluid, 
vs time, for various different system sizes, thus droplet 
sizes. Results from NVE as well as NVT (NHT) are presented. 
(b) Plot of ${D_d}{d_d}$ vs the droplet 
diameter, $d_d$, for the binary fluid. Results correspond 
to an averaging over $8$ independent initial configurations 
in NVE.}
\end{figure}
\par
\hspace{0.2cm}To substantiate the above difference 
further, we investigate if the number of particles in a 
droplet, before it collides, changes with time, for NHT. 
In Fig. \ref{fig6} we show $N_p$ in a typical large droplet 
as a function of time during the period when it does not 
undergo collision. The nature of the plot is rather flat, 
within statistical fluctuations. Notice that the period 
of time over which this plot is shown, the average number 
of particles in liquid droplets has significantly increased 
(see Fig. \ref{fig3}) due to collision among other 
droplets. This confirms that, in this case, 
the droplets do not grow due to deposition of additional 
particles onto them via standard diffusion mechanism, as 
in the LS case. In such a situation, the only way for the 
droplets to grow must be the collisions when they move.
\par
\hspace{0.2cm}
One possibility for the droplet motion is that the 
droplets are very compact objects and all the constituent 
particles move together. However, if this picture is true, 
keeping the conservation of total momentum in the 
system, motion of droplets will be constrained if we 
accept the temperature in the liquid and vapor domains to 
be same. But this we discard, following information in 
Fig. \ref{fig7} where we plot the fraction, $f_{ori}$, 
of original particles at time $t_0$, moving with a 
typical droplet as a function of time translation 
$t-t_0$. It appears that this quantity decays very fast, 
even though between collisions the size of the droplets 
remains unchanged. This essentially means that particles 
from different directions, possibly all, join the droplet 
to give (possibly) a random directional impact.
\par
\hspace{0.2cm}Often, in the molecular dynamics 
simulations in canonical ensemble, it is observed that 
application of thermostats brings in undesired features. 
It is, however, understood that a simulation in 
microcanonical (NVE) ensemble preserves all the 
requirements of hydrodynamics. But this ensemble, as 
already stated, is unwanted to study the nonequilibrium 
dynamics in the two-phase region. Needless to mention, 
it is a perfect candidate when one has only one 
droplet in the system, i.e., configuration-wise the 
system has achieved equilibrium. So, it will be a good 
idea to compare the results of Fig. \ref{fig7} with the 
equilibrium NVE MD runs. This, in addition to providing 
a comparative microscopic picture of droplet motion in 
equilibrium and nonequilibrium situations, will be a 
good test to confirm the validity of NHT in the 
study of hydrodynamic phenomena. 
\par
\hspace{0.2cm}In Fig. \ref{fig8}, we present the plots 
for $f_{ori}$, on a semi-log plot, for three different 
cases, viz., equilibrium in NVE, equilibrium with NHT 
and nonequilibirum with NHT. In all the cases, the 
droplet sizes are roughly the same. A very nice 
consistency of the curve for the nonequilibrium 
situation with NHT with the other ones speaks about the 
usefulness of NHT in such studies. On the other hand, 
linear looks of the curves on this semi-log plot is 
informative of an exponential decay of original particles 
from the droplet. Note that the system size used for 
the equilibrium cases is smaller than the nonequilibrium 
case. This is due to the fact that in equilibrium only 
one droplet exists and so a smaller system size is a 
necessity to match the size of the droplet with the 
nonequilibrium case. 
\par
\hspace{0.2cm}
In Fig. \ref{fig9}(a) we plot $f_{ori}$ vs $t-t_0$, 
for different droplet sizes from equilibrium 
configurations in NVE ensemble. It appears that the 
time scale of decay strongly depends on the size of 
the droplets. Note that the particles leave a droplet 
from the boundary or surface region. Considering the fact 
that the quantity of interest here is a volume fraction 
and further the length dependence of volume to surface 
ratio, it is expected that the time scale of decay should 
be longer for larger droplets. It appears that there is 
nice collapse of data from 
different droplet radii (see Fig. \ref{fig9}(b)), when 
the abscissa is scaled by the droplet surface area. 
According to the above argument, however, the time scale should 
have a linear dependence upon $d_d$. But, we could not 
yet construct an argument for this quadratic nature which 
appears to be the fact. Note that the dynamics is rather 
complex and one needs further probing to understand this 
behavior appropriately. Similar results 
from a binary Lennard-Jones fluid are also included in 
Fig. \ref{fig10}(a) and (b). Here the appropriate quantity 
is the fraction of original A-particles (the droplets 
contain A-particles as the majority species) moving with 
the droplet. The results are obtained from MD simulations 
in NVE ensemble with equilibrium configurations. Here 
also the qualitative nature of the decay is same as in 
the vapor-liquid case.
\par
\hspace{0.2cm}After the detour on the discussion of 
compactness of the droplets, we come back to investigate 
the nature of the droplet motion. In 
Fig. \ref{fig11}(a) and (b) we show a comparison between 
trajectories of typical droplets in the equilibrium and 
nonequilibrium situations from NVT (NHT). In both the 
cases, droplets are of same size and the trajectories 
are recorded over the same time length. Both the 
trajectories look random. However, it is clearly 
visible that the droplet 
in the nonequilibrium case spans larger distance than 
the equilibrium droplet. This is an indication 
of the presence of inter-droplet interaction in the 
nonequilibrium (multi-droplet) situation. To further 
strengthen our claim, in Fig. \ref{fig12} we plot 
the droplet MSD vs $t$ for two droplets of same size 
from equilibrium and nonequilibrium configurations 
using NHT. While the MSD in the equilibrium case 
shows linear behavior at late time, a characteristic 
of the Brownian diffusive motion, the nonequilibrium 
case does not show this feature. In the equilibrium 
case, there is only one droplet and so the inter-droplet 
interaction does not come into the picture, thus the 
motion is expected to be Brownian. Also, as already 
claimed, the MSD value for the nonequilibrium 
situation is higher than the equilibrium case. All 
these point to the presence of inter-droplet force. 
Note that in the nonequilibrium case, it is not possible 
to obtain MSD over an extended period of time because of 
the finite time available before the droplets collide. 
Of course, this would be possible in the dilute droplet 
density limit. But this will require very large 
system and we do not have possession of enough 
computational resources that this exercise would demand. 
We close the discussion on inter-droplet interaction in 
nonequilibrium situation by stating that our estimate 
\cite{roy1} for the amplitude ratios for the growths 
obtained using NHT and AT gives 
significantly higher value than the one quoted in 
Eq. (\ref{A_ratio}) for $A_{BS}/A_{LS}$. 
This also suggests the presence 
of this interaction.
\par
\hspace{0.2cm}
Next, to further justify the utility of NHT as a 
hydrodynamics preserving thermostat, we present results 
for the MSD in equilibrium from NHT as well as NVE, in 
Fig. \ref{fig13}(a). In addition to this justification, 
these results will provide important physics information 
as well. Results for different droplet 
sizes are plotted, as indicated in the figure. The 
dashed line here corresponds to the diffusive 
($\sim t$) behavior. Results from NHT are in excellent 
agreement with NVE. Here we should mention that in our 
MD simulations in NHT, the temperature was controlled 
by coupling with the absolute velocities, a standard 
practice followed. There are more recent algorithms 
where, instead of the absolute velocities, one uses 
the relative velocities that provides Galilean 
invariance \cite{schmid}. In future, it would be 
interesting to compare our simple NHT results with the 
latter to see what further improvement one can bring in.
\par
\hspace{0.2cm}Next we calculate the droplet 
diffusivity $D_d$, in equilibrium condition, for 
various different system sizes that automatically 
leads to different droplet sizes. In Fig. \ref{fig13}(b) 
we plot ${D_{_d}}{d_d}$ vs the droplet diameter $d_d$. 
Within statistical fluctuation, ${D_{_d}}{d_d}$ 
remains constant. This validates the SES relation 
even in such complex droplet diffusion case.
\par
\hspace{0.2cm}In Fig. \ref{fig14}(a) we plot the MSD 
of the equilibrium droplets in 
the binary fluid case for various different system 
sizes. All of them show the linear diffusive behavior, 
characteristic of the Brownian motion. Note that 
unlike the vapor-liquid case, the initial faster $t^2$ 
regime is absent here. This can be attributed to the 
difference in density of the background fluids in the 
two cases. In the vapor-liquid situation the low 
density of the background fluid facilitates the ballistic 
like faster motion of the droplet at early time leading 
to the $t^2$ regime, whereas in the binary fluid case 
the high density background suppresses it. For the 
same reason, overall MSD in the binary fluid is much less even 
though in both the cases results are presented for 
same time length. Here also we plot the MSD from both NVE and 
NVT (NHT) ensembles. The results for NVE are again nicely 
consistent with the NVT results. Exercises in 
Fig. \ref{fig13}(a) and \ref{fig14}(a) once more confirm 
the validity of NHT in studying hydrodynamic phenomena. 
In Fig. \ref{fig14}(b) we plot ${D_d}{d_d}$ vs $d_d$ for 
equilibrium binary fluid. The results are again in agreement 
with the SES relation.

\section{Conclusion}\label{conclusion}
\par
\hspace{0.2cm}In conclusion, we have studied the 
dynamics of droplets during the nonequilibrium evolution 
of a phase separating fluid \cite{suman3,roy1}, being 
in the nucleation regime of the phase diagram. 
Molecular dynamics 
simulations were used where a Nos\'{e}-Hoover thermostat 
controlled the temperature. The average volume of 
droplets grows linearly with time. It is demonstrated 
that the droplets grow via collisions. The motion 
of the droplets is not entirely random because of 
inter-droplet interaction even for reasonable low droplet 
density. This information was obtained from comparative 
study of the droplet dynamics in nonequilibrium situation 
and the equilibrium case. In the latter case, very similar 
results from the NVE ensemble molecular dynamics and those 
from the NVT ensemble with a Nos\'{e}-Hoover thermostat 
confirms the usefulness of this thermostat for the study 
of hydrodynamic phenomena. In equilibrium, the droplets 
exhibit Brownian motion and obey the 
Stokes-Einstein-Sutherland relation. Results for the equilibrium 
droplet dynamics are presented for a binary fluid 
\cite{das1,roy2} as well.
\par
\hspace{0.2cm}It is 
observed that the particles in a droplet do not move 
together even though between collisions the droplets 
keep their size intact. During the course of the motion, 
some particles leave and approximately a 
same number of particles join from various directions. 
We have found that the fraction of original particles moving 
with a droplet decays exponentially fast. 
\par
\hspace{0.2cm}In nonequilibrium situation 
it would be interesting to take a look at the space 
dependence of temperature. Of particular interest would be 
to study the temperature profile at the meeting area of 
two colliding droplets. It is expected that such collisions 
would give rise to some temperature increase. 
If so, understanding of how that enhanced temperature decays 
to the system value with time should also be of significant 
interest. Study with varying overall density will 
be of utmost importance to understand the dynamics further. 
In addition, understanding of decay of autocorrelation 
functions, probing the aging dynamics \cite{zannetti} 
in these systems, 
will be of general interest. 
\par
\hspace{0.2cm}
In this paper, the 
results for the nonequilibrium phenomena are presented only 
for the vapor-liquid phase separation. Needless to say, a 
similar study for a binary fluid will be very interesting. 
This latter problem will, of course, be computationally 
extremely demanding because of the overall high particle 
density throughout the system.

\section*{Acknowledgement}\label{acknowledgement}
SKD acknowledges discussion with K. Binder and F. 
Schmid. SKD and SR acknowledge financial support from the 
Department of Science and Technology, India, via 
Grant No SR/S2/RJN-$13/2009$. SR is grateful to the Council of 
Scientific and Industrial Research, India, for their research 
fellowship.


\vskip 0.5cm
\begin{thebibliography}{100}

\bibitem{zettlemoyer}\textit{Nucleation}, edited by A.C. Zettlemoyer (Dekker, 
New York, 1969).
\bibitem{abraham}F.F. Abraham, \textit{Homogeneous Nucleation Theory} 
(Academic, New York, 1974).
\bibitem{binder5}K. Binder, Rep. Prog. Phys. \textbf{50}, 783 (1987).
\bibitem{kashchiev}D. Kashchiev, \textit{Nucleation: Basic theory with 
Applications} (Butterworth-Heinemann, Oxford, 2000).
\bibitem{binder1}K. Binder in \textit{Kinetics of Phase Transitions} 
(CRC Press, Boca Raton, 2009), edited by S. Puri and V. Wadhawan.
\bibitem{gelb}L.D. Gelb, K.E. Gubbins, R. Radhakrishnan and M. Sliwinska-
Bartkowiak, Rep. Prog. Phys. \textbf{62}, 1573 (1999).
\bibitem{sing}\textit{Handbook of Porous Solids}, edited by 
F. Sch\"{u}th, K.S.W. Sing, and J. Weitkamp (Wiley-VCH, Weinheim, 2002).
\bibitem{squires1}T.M. Squires and S.R. Quake, Rev. Mod. Phys. 
\textbf{77}, 977 (2005).
\bibitem{bray}A.J. Bray, Adv. Phys. \textbf{51}, 481 (2002).
\bibitem{onuki}A. Onuki, \textit{Phase Transition Dynamics} 
(Cambridge University Press, UK, 2002).
\bibitem{binderbook}K. Binder, in \textit{Phase transformation of 
Materials}, edited by R.W. Cahn, P. Haasen and E.J. Kramer 
(VCH, Weinheim, 1991), Vol.\textbf{5}, p.405.
\bibitem{jones}R.A.L. Jones, \textit{Soft Condensed Matter} (Oxford 
University Press, Oxford, 2008).
\bibitem{lifshitz}I.M. Lifshitz and V.V. Slyozov, J. Phys. Chem. Solids 
\textbf{19}, 35 (1961).
\bibitem{suman1}S. Majumder and S.K. Das, Phys. Rev. E \textbf{81}, 050102 (2010).
\bibitem{suman2}S. Majumder and S.K. Das, Phys. Rev. E \textbf{84}, 021110 (2011).
\bibitem{hansen}J.-P. Hansen and I.R. McDonald, \textit{Theory of 
Simple Liquids} (Academic Press, London, 2008).
\bibitem{shaista1}S. Ahmad, S.K. Das and S. Puri, Phys. Rev. E. 
\textbf{82}, 040107 (2010).
\bibitem{suman3}S. Majumder and S.K. Das, Europhys. Lett. \textbf{95}, 
46002 (2011).
\bibitem{furukawa1}H. Furukawa, Phys. Rev. A \textbf{31}, 1103 (1985).
\bibitem{furukawa2}H. Furukawa, Phys. Rev. A \textbf{36}, 2288 (1987).
\bibitem{siggia}E.D. Siggia, Phys. Rev. A \textbf{20}, 595 (1979).
\bibitem{binder2}K. Binder and D. Stauffer, Phys. Rev. Lett. 
\textbf{33}, 1006 (1974).
\bibitem{binder3}K. Binder, Phys. Rev. B \textbf{15}, 4425 (1977).
\bibitem{tanaka1}H. Tanaka, J. Chem. Phys. \textbf{105}, 10099 (1996).
\bibitem{tanaka2}H. Tanaka, J. Chem. Phys. \textbf{107}, 3734 (1997).
\bibitem{tanaka3}H. Tanaka, J. Chem. Phys. \textbf{103} (6), 2361 (1995)
\bibitem{kumaran1}V. Kumaran, J. Chem. Phys. \textbf{109}, 7644 (1998).
\bibitem{kumaran2}V. Kumaran, J. Chem. Phys. \textbf{112}, 10984 (2000).
\bibitem{roy1}S. Roy and S.K. Das, Phys. Rev. E \textbf{85}, 050602 (2012).
\bibitem{perrot}F. Perrot, P. Guenoum, T. Baumberger and D. Beysens, 
Phys. Rev. Lett. \textbf{73}, 688 (1994).
\bibitem{squires2}T.M. Squires and J.F. Brady, Phys. Fluids \textbf{17}, 
073101 (2005).
\bibitem{das2}S.K. Das, J.V. Sengers and M.E. Fisher, J. Chem. Phys. 
\textbf{127}, 144506 (2007).
\bibitem{frenkel}D. Frenkel, B. Smit, \textit{Understanding Molecular 
Simulations: From Algorithm to Applications} (Academic Press, San Diego, 2002).
\bibitem{allen}M.P. Allen and D.J. Tildesley, \textit{Computer Simulations of Liquids}
(Clarendon, Oxford, 1987).
\bibitem{hoover}W.G. Hoover, \textit{Computational Statistical Mechanics}, 
Volume 11 of \textit{Studies in Modern Thermodynamics}, Elsevier (1991).
\bibitem{daspuri}S.K. Das and S. Puri, Phys. Rev. E. \textbf{65}, 026141 (2002).
\bibitem{das1}S.K. Das, J. Horbach, S. Puri, K. Binder, M.E. Fisher and 
J.V. Sengers, J. Chem. Phys. \textbf{125}, 024506 (2006).
\bibitem{roy2}S. Roy and S.K. Das, Europhys. Lett. \textbf{94}, 36001 (2011).
\bibitem{schmid}M.P. Allen and F. Schmid, Molecular Simulation \textbf{33}, 
21 (2006).
\bibitem{zannetti}M. Zannetti, in \textit{Kinetics of Phase Transitions} 
(CRC Press, Boca Raton, 2009), edited by S. Puri and V. Wadhawan.

\end{thebibliography}
\end{document}